\documentclass[twocolumn,aps,prl,superscriptaddress,showpacs,amsmath]{revtex4-1}
\usepackage{amsfonts}
\usepackage{amssymb}
\usepackage{mathrsfs}
\usepackage{graphicx}
\usepackage{float}
\usepackage{xcolor}
\usepackage{bm}
\usepackage{hyperref}
\usepackage{lineno}
\usepackage{multirow}
\usepackage{makecell}

\begin{document}
	
\title{Engineering exact mobility edges in quasiperiodic Aharonov-Bohm chains}

\author{Hai-Ying Cui}
\address{Innovation Academy for Precision Measurement Science and Technology, Chinese Academy of Sciences, Wuhan 430071, China}
\address{University of Chinese Academy of Sciences, Beijing 100049, China}

\author{Yi-Cong Yu}

\address{Innovation Academy for Precision Measurement Science and Technology, Chinese Academy of Sciences, Wuhan 430071, China}

\author{Xiaoming Cai}
\email{cxmpx@wipm.ac.cn}

\address{Innovation Academy for Precision Measurement Science and Technology, Chinese Academy of Sciences, Wuhan 430071, China}
\date{\today}

\begin{abstract}
	We investigate localization phenomena and exact mobility edges in a quasiperiodic Aharonov-Bohm chain, where the 1D canonical diagonal and off-diagonal Aubry-Andr\'{e}-Harper models are laterally coupled to an auxiliary sublattice threaded by a synthetic magnetic flux. By analytically computing the Lyapunov exponent via Avila's global theory of one-frequency Schr\"{o}dinger operators, we derive exact expressions for mobility edges. In the diagonal limit, the coupling to the auxiliary sublattice generates hyperbolic mobility edges that exhibit a sign-changing divergence and can be continuously tuned by the magnetic flux. In the off-diagonal regime, the interference between quasiperiodic hopping and indirect tunneling through the auxiliary sublattice gives rise to exact anomalous mobility edges that separate critical and extended states. Critical states and anomalous mobility edges emerge even in the absence of incommensurately distributed zeros in the hopping modulation, a behavior distinct from that of conventional off-diagonal models. Our results establish a rigorous theoretical framework for engineering controllable mobility edges, with the synthetic flux providing a tunable experimental knob that paves the way for realizations in platforms such as superconducting quantum circuits and photonic waveguides.
\end{abstract}
\maketitle

\section{I. Introduction}

The phenomenon of quantum localization, first established in Anderson's pioneering work, has remained a cornerstone of condensed matter physics for decades \cite{Anderson1958}. Disorder suppresses wave propagation and gives rise to spatially confined eigenstates. Scaling arguments dictate that in one- and two-dimensional systems, arbitrarily weak random disorder localizes all single-particle states exponentially \cite{Abrahams1979}. The situation becomes more intricate in three dimensions, where the energy spectrum can accommodate both extended and localized states, separated by a characteristic energy- the mobility edge (ME) - that marks the transition between them \cite{Evers2008,Lagendijk2029}. The existence of MEs underpins several fundamental physical phenomena, most notably the metal-insulator transition, and is also associated with pronounced thermoelectric effects that hold promise for energy conversion applications \cite{Whitney2014,Yamamoto2017,Chiaracane2020}.

Occupying a conceptual middle ground between completely random and perfectly periodic lattices, quasiperiodic structures exhibit unique properties rooted in their built-in spatial correlations. They can host localization phase transitions and MEs even in one dimension (1D). A paradigmatic example is the 1D Aubry-Andr\'{e}-Harper (AAH) model \cite{Aubry1980,Harper1955}, which displays a sharp transition from a global extended phase to a localized one at a finite strength of the quasiperiodic potential. When quasiperiodic modulations are further introduced in the off-diagonal hoppings, the resulting off-diagonal AAH model supports a critical phase \cite{Liu2015}. Critical states have recently attracted intensive research interest \cite{Yao2019a,Wang2020a,Xiao2021,Liu2024a,Roy2018,Jagannathan2021,Wang2025,Hopjan2023,Huang2025,Wnag2021b} owing to their distinctive properties, such as self-similarity, multifractality, and critical quantum dynamics. The localization in both canonical models is energy-independent, and consequently no ME emerges.

By incorporating additional ingredients-such as finite-range \cite{Roy2021b} or power-law decaying hoppings \cite{Deng2019}, synthetic spin-orbit interactions \cite{Zhou2013,Kohmoto2008}, tailored on-site potentials \cite{Biddle2009,Saha2019,Duthie2021,Dwiputra2022,Sarma1988,An2021}, or the mixing of clean and quasiperiodic chains \cite{Rossignolo2019,Lin2023,Lin2024} -it becomes possible to generate MEs within various extended AAH frameworks. In some specific quasiperiodic AAH models, exact formulas for MEs have been obtained through duality transformations \cite{Biddle2009,Ganeshan2015,Cedzich2024}, Avila's global theory \cite{Wang2020}, or the renormalization group method \cite{Gonçalves2023a,Gonçalves2023b}. Yet, only a handful of such generalized models exist \cite{Bodyfelt2014,Liu2022a,Wang2021,Hu2025,Zhang2025}. Despite their scarcity, these special cases offer invaluable analytical footholds for probing the physics of MEs, with implications that span noninteracting and interacting regimes-including studies of directed transport \cite{Whitney2014}, quantum thermal machines \cite{Chiaracane2020}, rectification of energy flow \cite{Balachandran2019}, superradiant phenomena \cite{Yin2020}, and many-body localization \cite{Kohlert2019}. The ability to derive exact analytical expressions for characteristic length scales and MEs in various AAH models plays a crucial role in advancing the fundamental understanding of localization transitions, providing theoretical insights that go well beyond numerical simulations.

While conventional MEs demarcate the boundary between extended and localized states, recent studies have unveiled a distinct class known as anomalous mobility edges (AMEs), which separate multifractal critical states from either localized or extended states \cite{Zhang2022,Duncan2024}. The emergence of AMEs significantly enriches the spectral landscape of quasiperiodic systems and provides deeper insights into the interplay between localization and criticality. To date, however, exactly solvable models exhibiting AMEs remain scarce \cite{Liu2022,Zhou2023,Li2025,Banerjee2025,Lu2025,Zhou2025}. Concurrently, advances originating from Avila's global theory \cite{Avila2015} have sharpened the rigorous characterization of critical states, indicating that such states typically arise in two scenarios: in diagonal AAH models, when the quasiperiodic potential is unbounded \cite{Banerjee2025}; in off-diagonal AAH models, when the  off-diagonal modulations possess incommensurately distributed zeros in the thermodynamic limit \cite{Liu2024}. This raises a natural and compelling question: do alternative scenarios or mechanisms exist that can induce critical states and AMEs in a controllable manner?


In this paper, we propose and solve a quasiperiodic Aharonov-Bohm (AB) chain that answers this question affirmatively. We laterally couple the 1D canonical diagonal and off-diagonal AAH models to an auxiliary sublattice threaded by a synthetic magnetic flux. This side-coupled AB geometry introduces a key physical ingredient: an indirect tunneling path between the primary lattice sites via the auxiliary sublattice. The interference between the direct quasiperiodic hopping and this indirect tunneling generates an effective hopping that can develop incommensurately distributed zeros even when the original hopping modulation is strictly positive. It is this interference-induced mechanism that gives rise to exact AMEs and critical states in our model.

By analytically computing the Lyapunov exponent (LE)-the inverse of localization length-via Avila's global theory of one-frequency Schr\"{o}dinger operators \cite{Avila2015}, we derive exact expressions for the (A)MEs that separate extended, critical, and localized states. In the diagonal limit, the coupling to the auxiliary sublattice generates hyperbolic MEs that exhibit a sign-changing divergence and can be continuously tuned by the magnetic flux. In the off-diagonal regime, the interference between quasiperiodic hopping and indirect tunneling through the auxiliary sublattice gives rise to AMEs that separate extended and critical states. Critical states and AMEs emerge even in the absence of incommensurately distributed zeros in the hopping modulation, a behavior distinct from the traditional off-diagonal scenario. Our results establish a rigorous theoretical framework for engineering controllable MEs in quasiperiodic chains, with the synthetic flux providing a tunable experimental knob that paves the way for realizations in platforms such as photonic waveguides \cite{Chang2025}, ultracold atoms \cite{Wang2022,Luschen2018}, cavity-polariton devices \cite{Goblot2020}, and superconducting quantum circuits \cite{Li2023}.

The remainder of this paper is organized as follows. In Sec. II, we introduce the quasiperiodic AB chain and present its Hamiltonian. Section III is devoted to the analytical derivation of exact MEs through the computation of the LE via Avila's global theory. In Sec. IV, we analyze the diagonal model and give the exact MEs that separate extended and localized states. Section V focuses on the off-diagonal model and the emergence of exact AMEs separating critical and extended/localized states. Finally, we present our conclusions in Sec. VI, together with an outlook on possible experimental realizations.

\section{II. Model and Hamiltonian}

\begin{figure}[tbp]
	\centering
	\includegraphics[scale=1, bb=0 0 230 66]{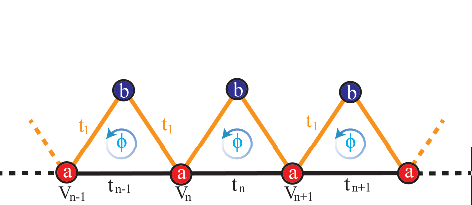}
	\caption{Schematic illustration of the quasiperiodic AB chain.}
	\label{Fig1}
\end{figure}

The quasiperiodic AB chain, as illustrated in Fig.\ref{Fig1}, is described by the following tight-binding Hamiltonian
\begin{eqnarray}
	H=&&\sum_{n}\left(t_nc^\dagger_{n,a}c_{n+1,a}+t_1e^{i\phi}c^{\dagger}_{n,a}c_{n,b}+t_1c^{\dagger}_{n,b}c_{n+1,a} +h.c\right) \nonumber\\
	&&+\sum_{n}\left(\Delta c_{n,b}^{\dagger}c_{n,b}+V_nc^{\dagger}_{n,a}c_{n,a}\right).
	\label{Hamiltonian}
\end{eqnarray}
Here $c^\dagger_{n,a/b}$ ($c_{n,a/b}$) are the creation (annihilation) operators on the $a/b$ sublattice sites of the $n$-th unit cell. $t_1$ is the hopping amplitude between $a$ and $b$ sublattices. $\Delta$ is a uniform on-site potential shift applied to the $b$ sublattice. The phase $\phi\in(-\pi,\pi]$ corresponds to the synthetic magnetic flux threading each triangle. We adopt a gauge where the flux induces a phase only for the intra-cell hopping between $a$ and $b$ sites. $t_n$ and $V_n$ are, respectively, the modulated hopping amplitude and on-site potential on $a$ sublattice, given by
\begin{eqnarray}
	t_n&=&t+W\cos\left[2\pi\beta\left(n+1/2\right)+\theta\right],\nonumber\\
	V_n&=&2V\cos\left(2\pi\beta n +\theta\right).
\end{eqnarray}
The parameter $t$ is the unmodulated part of the hopping amplitude. Without loss of generality, we will set $t=1$ as the unit of energy. $W\geqslant 0$ and $V\geqslant 0$ are the modulation amplitudes of the off-diagonal hopping and on-site potential, respectively. The global phase $\theta$ does not affect the localization properties, and we will set $\theta = 0$ unless otherwise noted. $\beta$ is an irrational number that characterizes the quasiperiodicity of modulations. 
The modulations are incommensurate with respect to the underlying lattice, act as quasirandom disorder, and result in the localization of states.
We take the value of inverse golden ratio [$\beta=(\sqrt{5}-1)/2$] as usual, and in numerical calculations it is approximated by rational numbers $\beta=F_k/F_{k+1}$, where $F_k$ is the $k$-th Fibonacci number.
Correspondingly, the total number of unit cells is chosen as $L=F_{k+1}$ to ensure periodic boundary conditions.

The Hamiltonian on the $a$ sublattice alone is identical to the canonical off-diagonal AAH model \cite{Liu2015}, which further reduces to the canonical diagonal AAH model \cite{Aubry1980} when $W=0$. When decoupled from the $b$ sublattice ($t_1=0$), the model is known to host distinct phases. For $0\leqslant W<t$, a metal-insulator transition occurs at $V=t$: all states are extended when $V<t$ and localized when $V>t$. For $W\geqslant t$, the system undergoes the localization transition at $V=W$ from a critical phase, where all states are multifractal, to the localized phase for $V>W$. In the other limit $t_n=0$, the model maps onto a mosaic AAH model, which is known to support exact MEs at $|E|=V^{-1}$ \cite{Wang2020}. States with energies between these two MEs are extended, while outside are localized. The aim of this work is to show how coupling these ingredients via the AB cage engineers and controls exact MEs.

\section{III. Computing Lyapunov exponent}

In this section, we compute the LEs of eigenstates. For a single-particle state $|\Psi\rangle=\sum_n(\psi^a_nc^\dagger_{n,a}+\psi^b_nc^\dagger_{n,b})|0\rangle$ with eigenenergy $E$, the eigenequations are
\begin{eqnarray}
	&&t_n\psi^a_{n+1}+t_{n-1}\psi^a_{n-1}+t_1e^{i\phi}\psi^b_n+t_1\psi^b_{n-1}+V_n\psi^a_{n}=E\psi^a_{n},\nonumber\\ &&t_1\psi^a_{n+1}+t_1e^{-i\phi}\psi^a_{n}+\Delta\psi^b_{n}=E\psi^b_{n}.
	\label{EigenEq}
\end{eqnarray}
Substituting the second equation into the first yields an effective eigenequation for amplitudes $\psi^a_{n}$:
\begin{eqnarray}\label{EigenEq1}
	[t_n+\frac{t^2_1e^{i\phi}}{E-\Delta}]\psi^a_{n+1}+[t_{n-1}+&&\frac{t^2_1e^{-i\phi}}{E-\Delta}]\psi^a_{n-1}\\
	&&=[E-V_n-\frac{2t^2_1}{E-\Delta}]\psi^a_{n}.\nonumber
\end{eqnarray}
Notice that when $E=\Delta$, one can easily obtain directly from Eq.(\ref{EigenEq}) that the state is extended, the same as nearby states. In transfer matrix form, the effective eigenequation reads
\begin{equation}
	\left[\begin{array}{c}
		\psi^a_{n+1}\\
		\psi^a_n\end{array}
	\right]=T_n\left[\begin{array}{c}
		\psi^a_{n}\\
		\psi^a_{n-1}\end{array}
	\right],T_n=C_nD_n,
	\label{TM}
\end{equation}
with
\begin{eqnarray}
	C_n&=&(t_n+\frac{t_1^2e^{i\phi}}{E-\Delta})^{-1},\nonumber\\
	D_n&=&\left[\begin{array}{cc}
		-\frac{2t_1^2}{E-\Delta}-(V_n-E)&-t_{n-1}-\frac{t_1^2e^{-i\phi}}{E-\Delta}\\
		t_n+\frac{t_1^2e^{i\phi}}{E-\Delta}&0\end{array}
	\right].\nonumber
\end{eqnarray}
Since the transfer matrix factorizes as the commutative product of $C_n$ and $D_n$, the LE of state can be computed by
\begin{equation}
	\gamma=\lim_{L\rightarrow\infty}\frac{1}{L}\mathrm{ln}\|\prod_{n=1}^LT_n\|=\gamma_C+\gamma_D\nonumber
\end{equation}
where $\|\cdot\|$ denotes the matrix norm, which is defined by the largest absolute value of its eigenvalues. Using Weyl's equidistribution theorem for irrational $\beta$ \cite{Choe1993,Avila2015}, the contribution $\gamma_C$ is evaluated as
\begin{eqnarray}
	\gamma_C&=&\lim_{L\rightarrow\infty}\frac{1}{L}\mathrm{ln}|\prod_{n=1}^L(t_n+\frac{t_1^2e^{i\phi}}{E-\Delta})^{-1}|\nonumber\\
	&=&-\frac{1}{2\pi}\int\limits_{0}^{2\pi}\ln|t+W\cos(\xi)+\frac{t_1^2e^{i\phi}}{E-\Delta}|\mathrm{d}\xi
	\label{gammaC}
\end{eqnarray}
For the remaining matrix, we apply Avila's global theory of one-frequency analytic cocycles $\mathrm{SL}(2,\mathbb{C})$\cite{Avila2015,Wang2020}. The first step is to analytically continue the global phase $\theta$ in $D_n$ as  $\theta\rightarrow\theta+i\epsilon$. In the limit of $\epsilon\rightarrow\pm\infty$, a straightforward calculation gives
\begin{eqnarray}
	D_n(\epsilon\rightarrow\pm\infty)=e^{\pm(\epsilon-i2\pi\beta n- i\theta)}\left[\begin{matrix}
		-V & -We^{\mp i\pi\beta}/2\\
		We^{ \pm i\pi\beta}/2&0
	\end{matrix}
	\right],\nonumber
\end{eqnarray}
which leads to
\begin{eqnarray}
	\gamma_D(\epsilon\rightarrow\pm\infty)=|\epsilon|+\max\left\{\ln|(V\pm\sqrt{V^2-W^2})/2|,0\right\}.\nonumber
\end{eqnarray}
According to Avila's global theory \cite{Avila2015}, as a function of $\varepsilon$, $\gamma_D(\varepsilon)$ is a convex, piecewise linear function with integer slopes. Moreover, the theory shows that $\gamma_D(\varepsilon)$ is an affine function in the neighborhood of $\varepsilon=0$. Therefore, it suffices to compute the asymptotic behavior in the limits $\varepsilon\rightarrow\pm\infty$, and then use the convexity and integrality of the slopes to determine $\gamma_D(\varepsilon)$ at $\epsilon=0$. This yields
\begin{eqnarray}
	\gamma_D=\max\left\{\ln|(V\pm\sqrt{V^2-W^2})/2|,0\right\},
	\label{gammaD}
\end{eqnarray}
Combining the two contributions, the LE of state is
\begin{eqnarray}
\gamma=\left\{\begin{array}{l}\max(f_1+\gamma_C,0),\quad V\leqslant W,\\ \max(f_2+\gamma_C,0),\quad V>W,\end{array}\right.
\label{LE}
\end{eqnarray}
with
\begin{eqnarray}
	f_1=\ln |W/2|,\quad f_2=\ln|(V+\sqrt{V^2-W^2})/2|.
	\label{f12}
\end{eqnarray}
When $\gamma>0$, the state is Anderson localized, decaying asymptotically as $\mathrm{exp}(-\gamma|n-n_0|)$ around a localization center $n_0$. In contrast, $\gamma=0$ signals an extended or critical state. Therefore, the exact ME, marking the transition of state from extended or critical to localized, is determined by the condition $f_{1}+\gamma_C=0$ for $V\leqslant W$ and $f_{2}+\gamma_C=0$ for $V> W$. The integral in $\gamma_C$ [Eq.(\ref{gammaC})] can be evaluated using Jensen's formula \cite{Jen}, with the result depending on the parameter regime. Accordingly, the localization properties are classified into different regions, which we will discuss separately in the following two sections. We further note that the potential shift $\Delta$ on $b$ sublattice merely shifts the energy in the LE and consequently the MEs; for simplicity, we set $\Delta=0$ in the rest of this work. Thus, the remaining control parameters of the system are only $t_1$, $W$, $V$, and $\phi$. Moreover, the LEs $\gamma_C$ and $\gamma$ are even functions of $\phi$. They are also invariant under the simultaneous replacements $\phi\rightarrow\pi-\phi$ and $E\rightarrow-E$. Therefore, we can further restrict $\phi$ to the interval $[0,\pi/2]$.

\section{IV. Exact mobility edges in diagonal AB chain ($W=0$)}

\begin{figure}[tbp]
	\centering
	\includegraphics[scale=1, bb=0 0 277 176]{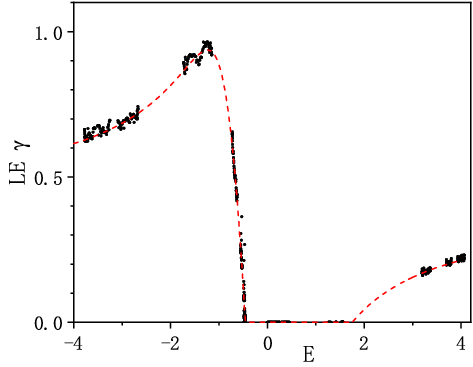}
	\caption{LEs of all single-particle states, as a function of energy, for the system with $L=610$, $W=0$, $t_1=1$, $\phi=0.2\pi$, and $V=1.5$. The red dashed line corresponds to Eq.(\ref{LE0}).}
	\label{FigAdd1}
\end{figure}

\begin{figure}[tbp]
	\centering
	\includegraphics[scale=1.1, bb=0 0 277 336]{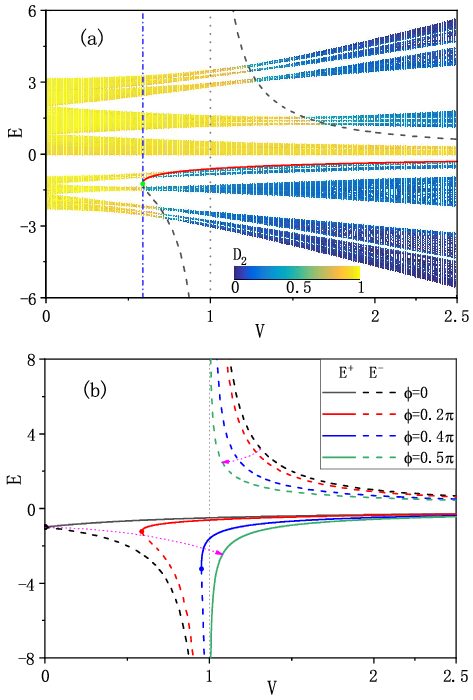}
	\caption{(a) Eigenenergies $E$ as functions of the quasiperiodic potential strength $V$ for diagonal AB chains with $L=610$, $W=0$, $t_1=1$, and $\phi=0.2\pi$. The fractal dimension $D_2$ of each state is color-coded. Gray dotted and blue dash-dotted lines mark the divergence ($V=t$) and the coincidence ($V=t\sin\phi$), respectively. (b) Exact MEs on the $(E,V)$ plane for systems with $W=0$, $t_1=1$, and various $\phi$. The gray dotted line marks the divergence at $V=t$.}
	\label{Fig2}
\end{figure}

In this section we focus on the situation with only the diagonal quasiperiodic potential. In this case, the model on $a$ sublattice is precisely the canonical AAH model, which hosts a global extended-localized transition at $V=t$. Introducing the coupling to $b$ sublattice can yield exact MEs that separate extended and localized states. Mathematically, when $W=0$, the integral kernel in $\gamma_C$ [Eq.(\ref{gammaC})] becomes $\xi$-independent. Then, the LE of state can be easily calculated as
\begin{eqnarray}
	\gamma=\max\left(\ln\left|\frac{VE}{tE+t_1^2e^{i\phi}}\right|,0\right).
	\label{LE0}
\end{eqnarray}
The condition $|VE/(tE+t_1^2e^{i\phi})|=1$ determines the exact MEs, which are located at the critical energies
\begin{eqnarray}
	E^{\pm}=-\frac{t_1^2}{t\cos\phi\pm\sqrt{V^2-t^2\sin^2\phi}}.
	\label{ME1}
\end{eqnarray}
These two critical energies are physically meaningful only when $V\geqslant t\sin\phi$. Thus, MEs exist solely for $V\geqslant t\sin\phi$, whereas for $V< t\sin\phi$ the system is in the extended phase, with $f_2+\gamma_C<0$ and $\gamma=0$ for all states. MEs are illustrated by lines in Fig.\ref{Fig2}(a). Both MEs satisfy $E^{\pm}\rightarrow0$ as $V\rightarrow\infty$. Furthermore, the critical energy $E^-$ diverges at $V=t$. More precisely, for $\phi\in[0,\pi/2)$, it tends to $-\infty$ as $V\rightarrow t^-$ and to $+\infty$ as $V\rightarrow t^+$. In contrast, the critical energy $E^+$ remains continuous at $V=t$ when $\phi\in[0,\pi/2)$. In addition, the critical energies $E^{\pm}$ coincide at $V=t\sin\phi$, for $\phi\neq \pi/2$.

The above analytical results agree well with numerical simulations. 
To validate the theoretically computed LE in Eq.(\ref{LE0}), we first fit single-particle eigenstates numerically, using exponential wave functions $\psi^{k,a/b}_n\propto\mathrm{exp}(-\gamma_k|n-n_0|)$. 
$n_0$ denotes the localization center, $k$ is the index of states, and $\gamma_k$ is the LE of state.
We present in Fig.\ref{FigAdd1} the LEs of all single-particle states, as a function of eigenenergy, for a typical system. In the middle of spectrum states are extended with zero LEs, while states are localized with finite LEs on two sides. Most importantly, the numerical LEs (black dots) agree well with the theoretical prediction in Eq.(\ref{LE0}) (red dashed line), considering the finite-size effect.

To further probe the localization properties in detail, we numerically compute the inverse participation ratios and fractal dimensions of states. 
The inverse participation ratio is defined as $P=\sum_n(|\psi^a_n|^4+|\psi^b_n|^4)$ for a normalized single-particle state. We find that it exhibits the same behavior as those defined on each sublattice separately. 
Generally, the inverse participation ratio $P \propto L^{-D_2}$, where $D_2$ is the fractal dimension. 
For an extended state, $P \propto 1/L$ and $D_2=1$, whereas the inverse participation ratio approaches unity and $D_2=0$ for a localized state. 
Critical states, characterized by $0<D_2<1$, exhibit multifractal properties \cite{Kutlin2024}. 
In Fig.\ref{Fig2}(a), we present typical eigenenergies $E$ versus the quasiperiodic potential strength $V$, with the $D_2$ values of all states color-coded to indicate their localization nature. Corresponding exact MEs are added. The critical energies $E^+$ and $E^-$ are presented as red solid and black dashed lines, respectively. In addition, the lines $V=t$ (gray dotted) and $V=t\sin\phi$ (blue dash-dotted) mark where the divergence and the coincidence occur. One can clearly see that the MEs match the numerical data well and properly separate extended and localized states. For $V<t\sin\phi$ the system is fully extended. However, when $\phi=0$, two MEs coincide at $V=0$ and the extended phase disappears. MEs emerge for $V\geqslant t\sin\phi$. Due to the divergence at $V=t$, when $t\sin\phi<V<t$ the states between two MEs are localized while those outside are extended, whereas for $V>t$ the situation is reversed: the states between MEs are extended and those outside are localized. Finally, we note that the localization map and MEs for the flux $\pi-\phi$ can be obtained from those for $\phi$ simply by reflecting about the $E=0$ axis.

At the end of this section, we analyze how the MEs evolve on the $(E,V)$ plane as parameters change. From Eq.(\ref{ME1}) it is clear that the hopping $t_1$ merely renormalizes the values of MEs, and we therefore focus on their $\phi$-dependence. In Fig.\ref{Fig2}(b), we plot MEs $E^+$ (solid lines) and $E^-$ (dashed lines) as functions of the quasiperiodic potential strength $V$, using different colors to represent distinct values of the magnetic flux $\phi$. The MEs display hyperbolic-like dispersions. The ME $E^+$ is always negative and increases towards $0$ from below as $V$ grows, while the ME $E^-$ exhibits a sign-changing divergence at the critical point $V=t$. When $\phi=0$, two MEs coincide at $(V=0,E=-1)$, located on the lower half plane, and no extended phase exists. As $\phi$ increases to $\pi/2$, this coincidence point shifts to the lower right and approaches $(V=t,E=-\infty)$, always remaining on the lower half plane. The extended phase emerges and gradually expands. On the lower half plane the MEs shift rightwards, but portions of them stay to the left of the divergence line $V=t$; on the upper half plane, the MEs shift leftwards but remain entirely to the right of the divergence line. When $\phi=\pi/2$, the critical energies become $E^{\pm}=\mp t^2_1/\sqrt{V^2-t^2}$, with $E=0$ and $V=t$ as the horizontal and vertical asymptotes, respectively. Two MEs are symmetric under reflection about $E=0$ and both lie to the right of the divergence line.

\section{V. Exact anomalous mobility edges in off-diagonal AB chain ($W\neq0$)}

When $W\neq0$, the Hamiltonian on $a$ sublattice is the canonical off-diagonal AAH model studied in Ref.\cite{Liu2015}, which hosts a global critical phase for $W\geqslant t,V$. Upon coupling to the $b$ sublattice, anomalous mobility edges (AMEs)-boundaries separating critical and extended states-emerge. Mathematically, for $W\neq0$ the integral kernel in $\gamma_C$ [Eq.(\ref{gammaC})] depends on $\xi$. Using Jensen's formula \cite{Jen}, we obtain
\begin{eqnarray}
	\gamma_C&=&\min(\ln|z_+|,\ln|z_-|)-\ln |W/2|,\nonumber\\
	z_\pm&=&-\eta\pm\sqrt{\eta^2-1},\nonumber\\
	\eta&=&t/W+t_1^2e^{i\phi}/WE.
	\label{eta}
\end{eqnarray}
Here $z_\pm$ are the two roots of the equation $z^2+2\eta z+1=0$. When $\phi=0$, $\eta$ is real, and two roots can lie on the unit circle, giving rise to critical states. We treat this case separately.

\subsection{A. Case $\phi=0$}

\begin{figure}[tbp]
	\centering
	\includegraphics[scale=1, bb=0 0 277 350]{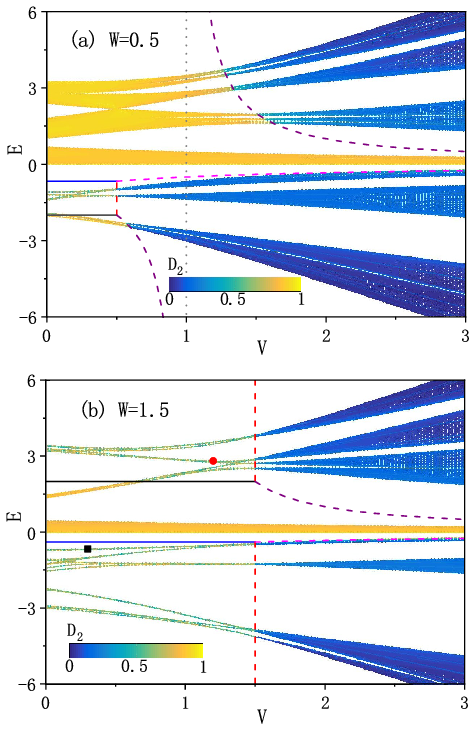}\\
	\hspace{-0.5cm}\includegraphics[scale=1.06, bb=0 0 227 178]{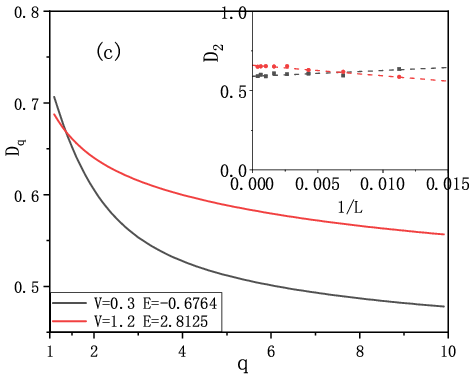}
	\caption{(a,b) Eigenenergies $E$ versus the quasiperiodic potential strength $V$, with $D_2$ values of all states color-coded, for off-diagonal AB chains with $L=610$, $t_1=1$, and $\phi=0$. (a) $W=0.5<t$. (b) $W=1.5>t$. Blue and black solid lines indicate the AMEs $E^+_1$ and $E^-_1$, respectively; pink and purple dashed lines indicate the MEs $E^-_2$ and $E^+_2$. The red dashed line is $V=W$, and the gray dotted line signals the divergence at $V=t$. (c) Generalized fractal dimension $D_q$ versus $q$ for two critical states (indicated by dots in (b)). Inset: Their finite-size scaling of $D_2$.}
	\label{Fig3}
\end{figure}

For $\phi=0$, $\eta=t/W+t_1^2/WE$ is real. Both roots satisfy $|z_\pm|=1$ when $|\eta|\leqslant 1$, yielding $\gamma_C=-\ln |W/2|$. The condition $|\eta|\leqslant 1$ corresponds to $E\in[E^-_1,E^+_1]$ for $W<t$, and to $E\in(-\infty,E^+_1]\cup[E^-_1,+\infty)$ for $W\geqslant t$, where we have defined $E^\pm_1=-t_1^2/(t\pm W)$. When $|\eta|>1$, two roots are real and reciprocal, with one root having modulus larger than $1$ and the other smaller than $1$. The condition $|\eta|>1$ gives $E\in(-\infty,E^-_1)\cup(E^+_1,+\infty)$ for $W<t$, and $E\in(E^+_1,E^-_1)$ for $W\geqslant t$. Incorporating the contribution from $\gamma_D$ given by Eqs.(\ref{gammaD}-\ref{f12}), which distinguishes two regimes, the localization properties fall into four parameter regions, as summarized in Table \ref{T1} and detailed below:

\noindent\textbf{i) $W<t$ and $V\leqslant W$:} For $E\in[E^-_1,E^+_1]$, $|z_\pm|=1$, $\gamma_C=-\ln |W/2|$, and $f_1+\gamma_C$ in Eq.(\ref{LE}) is always zero, implying $\gamma=0$ and critical states. For $E\in(-\infty,E^-_1)\cup(E^+_1,+\infty)$, two roots $z_\pm$ are real, and depending on the energy interval the root with smaller modulus is either $z_+$ or $z_-$; nevertheless, $f_1+\gamma_C$ is always negative, so $\gamma=0$ and states are extended. The exact AMEs are located at the negative energies
\begin{equation}
E^\pm_1=-t_1^2/(t\pm W),
\end{equation}
which separate extended and critical states. In Fig.\ref{Fig3}(a), we present typical eigenenergies $E$ versus $V$ for systems with $W<t$, with the fractal dimensions $D_2$ of states color-coded. The region $V\leqslant W$ lies to the left of the vertical red dashed line $V=W$. Exact AMEs $E_1^-<E_1^+$ are shown as horizontal black and blue lines, respectively, and they agree well with the numerical simulation. Critical states occupy the central part of the spectrum. These two AMEs are independent of $V$. The parameter $t_1$ rescales the AME energies. Thus, as $t_1$ increases the energy window of critical states widens. As $W$ grows, the AME $E^+_1$ moves upwards, while $E^-_1$ shifts downwards. Furthermore, the vertical red dashed line $V=W$ shifts to the right, thereby enlarging the critical region on the $(E,V)$ plane. Hence, both $t_1$ and $W$ favor the critical states.

\noindent\textbf{ii) $W<t$ and $V>W$:} In this regime the exact MEs are given by 
\begin{equation}
E_2^\pm=t_1^2/(\pm V-t).
\end{equation}
Corresponding detailed analytical derivation is deferred to the Appendix. The ME $E_2^-$ is always negative [pink dashed line in Fig.\ref{Fig3}(a)] and connects to the AME $E_1^+$ at $V=W$. As $V$ increases from $W$, $E_2^-$ grows monotonically towards zero. On the other hand, the ME $E_2^+=t_1^2/(V-t)$ connects to the AME $E_1^-$ at $V=W$. It displays a hyperbolic dispersion and undergoes a sign-changing divergence at $V=t$: it tends to $-\infty$ as $V\rightarrow t^-$ and to $+\infty$ as $V\rightarrow t^+$. Because of this divergence, states between (outside) two MEs are localized (extended) for $W<V<t$, whereas for $V>t$ the situation is reversed [see the right of vertical red dashed line in Fig.\ref{Fig3}(a)]. $t_1$ still rescales two MEs, but its effect differs on the two sides of the divergence: for $V<t$, increasing $t_1$ enlarges the energy window of localized states, while for $V>t$, it enlarges that of extended states. Two MEs are independent of $W$. However, as $W$ increases, the vertical red dashed line $V=W$ moves to the right, which shrinks the region $W<V<t$.

\noindent\textbf{iii) $W>t$ and $V\leqslant W$:} For $E\in(-\infty,E^+_1)\cup(E^-_1,+\infty)$, we have $|z_\pm|=1$, $\gamma_C=-\ln |W/2|$, and $f_1+\gamma_C$ in Eq.(\ref{LE}) is always zero; hence $\gamma=0$ and states are critical. For $E\in[E^+_1,E^-_1]$, $f_1+\gamma_C<0$, so $\gamma=0$ and states are extended. Two exact AMEs are located at critical energies $E^\pm_1=-t_1^2/(t\pm W)$, the same formulas as in case i). Note, however, that $E^-_1$ is now positive while $E^+_1$ remains negative. Moreover, in contrast to case i), states between (outside) two AMEs are extended (critical). In addition, increasing $t_1$ enlarges the energy window of extended states, while increasing $W$ shrinks it. See the left of vertical red dashed line in Fig.\ref{Fig3}(b) for a numerical simulation.

\noindent\textbf{iv) $W>t$ and $V>W$:} Exact MEs are again at critical energies $E_2^\pm=t_1^2/(\pm V-t)$, as in case ii), but the LE and localization properties differ (see the Appendix for details). Now $E_2^+$ is always positive, while $E_2^-$ is always negative; they exhibit no divergence. Two MEs connect to AMEs $E_1^\mp$ at $V=W$, and monotonically approach zero as $V$ increases from $W$ to $+\infty$. States between them are extended, while those outside are localized. See the right of vertical red dashed line in Fig.\ref{Fig3}(b). The MEs are independent of $W$, but the boundary $V=W$ shifts to the right with increasing $W$.

\begin{table*}[tp]
	\centering
	\begin{tabular}{|c|c|c|c|}
		\hline
		& $V\leqslant W$ & \multicolumn{2}{c|}{$V>W$}\\
		\hline	
		\multirow{2}{*}{Mobility edges}& \multirow{2}{*}{$E_1^\pm=-\frac{t_1^2}{t\pm W}$}&\multicolumn{2}{c|}{\multirow{2}{*}{$E_2^\pm=\frac{t_1^2}{\pm V-t}$}}\\
		& &\multicolumn{2}{c|}{}\\
		\hline
		Localization&Cri:$[E^-_1,E^+_1]$& $W<V<t$&$V>t$\\
		\cline{3-4}	
		$W<t$&Ext:$(-\infty,E^-_1)\cup(E^+_1,+\infty)$& Ext:$(-\infty,E^+_2)\cup(E^-_2,+\infty)$&Ext:$(E_2^-,E_2^+)$\\
		& &Loc:$(E_2^+,E_2^-)$&Loc:$(-\infty,E^-_2)\cup(E^+_2,+\infty)$\\
		\hline
		Localization&Cri:$(-\infty,E^+_1)\cup(E^-_1,+\infty)$&\multicolumn{2}{c|}{Ext:$(E_2^-,E_2^+)$}\\
		$W\geqslant t$&Ext:$(E^+_1,E^-_1)$&\multicolumn{2}{c|}{Loc:$(-\infty,E^-_2)\cup(E^+_2,+\infty)$}\\
		\hline		
	\end{tabular}\\
	
	\vspace{0.5em}
	Ext=Extended states; \quad Loc=Localized states; \quad  Cri=Critical states.	
	\caption{Exact (A)MEs and localization of states for the off-diagonal AB chain with $W>0$ and $\phi=0$.}
	\label{T1}
\end{table*}

To confirm the multifractal nature of the critical states, we compute the generalized inverse participation ratio $P^q=\sum_n(|\psi^a_n|^{2q}+|\psi^b_n|^{2q})\propto L^{(1-q)D_q}$ for $q>0$, where $D_q$ is the generalized fractal dimension. In Fig.\ref{Fig3}(c), we present $D_q$ versus $q$ for two typical critical states (marked as dots in Fig.\ref{Fig3} (b)). $0<D_q<1$ and $D_q$ is a monotonically decreasing function of $q$, which are two characteristic features of a multifractal critical state \cite{Kutlin2024}. The inset shows the finite-size scaling of $D_2$ for different system sizes with linear fittings indicated by dashed lines, confirming that the multifractal behavior persists in the thermodynamic limit.

Finally, we discuss the origin of critical states in our model. The canonical off-diagonal AAH model on $a$ sublattice alone is well known to host a critical phase when $W>t$. In that model, criticality has been linked to the presence of incommensurately distributed zeros in the hopping modulation \cite{Liu2024}. When $W>t$, the hopping amplitude $t_n=t+W\cos\left[2\pi\beta\left(n+1/2\right)\right]$ vanishes at a set of positions that are incommensurate with the lattice, thereby dividing the chain into segments of incommensurate lengths with nearly vanishing inter-segment couplings. The wave functions on these segments become quasiperiodically self-similar, giving rise to critical states. In our AB chain, however, even when $t_n$ possesses such zeros, neighboring $a$ sites (or segments) remain indirectly coupled through a common $b$ site. Therefore, the critical states we observe are not directly in the conventional framework of incommensurately distributed zeros. From Eqs.(\ref{EigenEq}-\ref{EigenEq1}), we ascribe their appearance to a destructive interference between the quasiperiodic hopping and the indirect tunneling through $b$ sublattice. This interference can cause the effective hopping entering the eigenequation for the amplitudes $\psi^a$ [Eq.(\ref{EigenEq1})] to develop incommensurately distributed zeros when $\phi=0$. Concretely, the effective hopping amplitude $t'_n=t+W\cos(2\pi\beta n)+t^2_1/(E-\Delta)$ possesses such zeros when $W\geqslant |t^2_1/(E-\Delta)+t|$, which corresponds precisely to the occurrence of critical states. The condition $W=|t^2_1/(E-\Delta)+t|$ yields exactly the AMEs $E^\pm_1$. The fact that critical states and AMEs exist only at $\phi=0$, as confirmed by numerical simulations, further solidifies this interpretation.

\subsection{B. Case $\phi\neq0$}

\begin{figure*}[tbp]
	\centering
	\includegraphics[scale=1.1, bb=0 0 454 327]{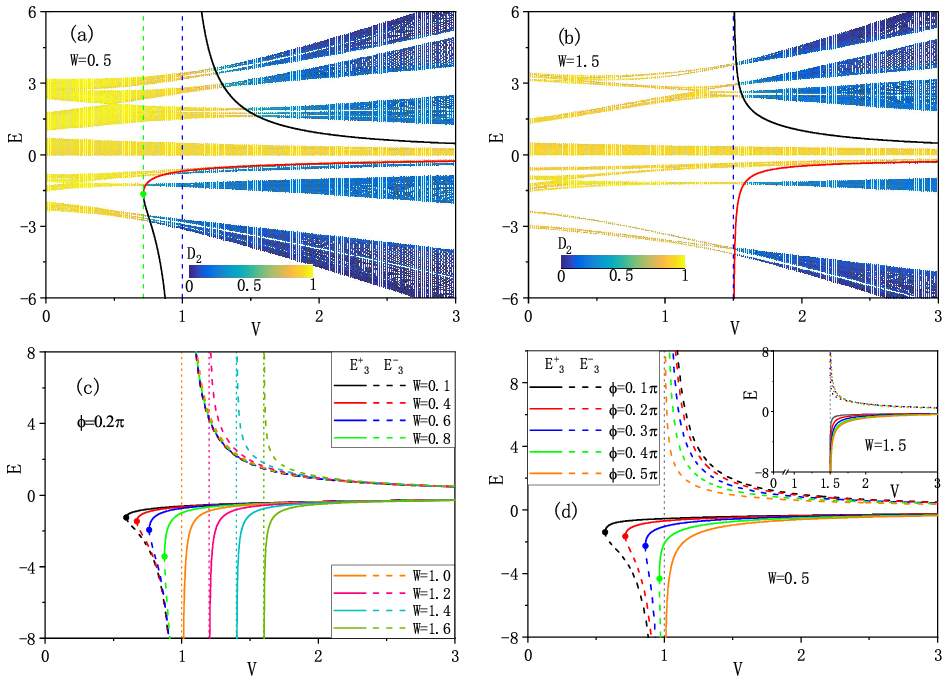}
	\caption{Eigenenergies $E$ as functions of $V$ for off-diagonal AB chains with $L=610$, $t_1=1$, and $\phi=0.2\pi$. The $D_2$ values are color-coded. (a) $W=0.5<t$. (b) $W=1.5>t$. Black and red solid lines correspond to the MEs $E^-_3$ and $E^+_3$, and blue and green dashed lines mark the divergence ($V=t$) and the coincidence ($V=V_c$), respectively. (c) Exact MEs on the $(E,V)$ plane for $t_1=1$, $\phi=0.2\pi$, and several values of $W$. Colored dotted lines indicate the asymptotes $V=W$ for the corresponding MEs. (d) Main panel and inset: Exact MEs on the $(E,V)$ plane for $t_1=1$ and different $\phi$, with $W=0.5$ (main) and $W=1.5$ (inset). The gray dotted line marks the divergence at $V=t$.}
	\label{Fig4}
\end{figure*}

For the general case $\phi\neq0$, the quantity $\eta$ in Eq.(\ref{eta}) is complex, and $|z_+|\neq |z_-|$. It is straightforward to verify that $|z_-|>|z_+|$ when $\mathrm{Re}(\eta)>0$, and $|z_-|<|z_+|$ when $\mathrm{Re}(\eta)\leqslant0$. The condition $\mathrm{Re}(\eta)>0$ corresponds to $E\in(-\infty,-t_1^2\cos(\phi)/t)\cup(0,+\infty)$ and yields $\gamma_C=\ln|2z_+/W|$, while $\mathrm{Re}(\eta)\leqslant 0$ corresponds to $E\in[-t_1^2\cos(\phi)/t,0)$ with $\gamma_C=\ln|2z_-/W|$. Including the contribution from $\gamma_D$ given in Eqs.(\ref{gammaD}-\ref{f12}), one can first prove that when $V\leqslant W$, $\gamma_C+f_1$ is always negative, leading to $\gamma=0$ and thus extended states [see Appendix]. For $V>W$, the LE takes the form $\gamma=\max(\ln|z_+S/W|,0)$ if $E\in(-\infty,-t_1^2\cos(\phi)/t)\cup(0,+\infty)$, and $\gamma=\max(\ln|z_-S/W|,0)$ if $E\in[-t_1^2\cos(\phi)/t,0]$, where $S$ is defined below in Eq.(\ref{ME3}). Although the LE has different expressions in two energy intervals, a tedious but straightforward calculation [detailed in the Appendix] shows that they lead to the same exact MEs
\begin{eqnarray}	
	E_3^\pm&=&\frac{-t_1^2\left[\left(\frac{S^2+W^2}{S^2-W^2}\right)^2\sin^2\phi+\cos^2\phi\right]}{t\cos\phi\pm V\sqrt{1-\frac{4S^2(t^2-W^2)}{(S^2-W^2)^2}\sin^2\phi}},\label{ME3}\\
	\nonumber\\
	S&=&V+\sqrt{V^2-W^2}>V>W.\nonumber
\end{eqnarray}
These two MEs are physically meaningful only when the expression under the square root is  non-negative. This imposes the condition $V\geqslant V_c>W$ when $W<t$, where the exact expression for $V_c$ is given in the Appendix. For $W<t$, two MEs exhibit a hyperbolic-like structure similar to the $W=0$ case in Eq.(\ref{ME1}). $E^+_3$ is always negative and increases monotonically towards zero as $V\rightarrow+\infty$ [see the lines in Fig.\ref{Fig4}(a)]. In contrast, $E^-_3$ undergoes a sign-changing divergence at $V=t$ for $\phi\in(0,\pi/2)$. Two MEs $E^\pm_3$ coincide at $V=V_c$. On the other hand, when $W\geqslant t$, MEs exist only for $V>W$ [see Fig.\ref{Fig4}(b)]. Here $E^+_3$ is negative and $E^-_3$ is positive; both diverge as $V\rightarrow W^+$ with asymptotic behavior $E^\pm_3\rightarrow \mp\infty$. As $V$ increases from $W$ to $+\infty$, $E^+_3$ ($E^-_3$) monotonically increases (decreases) toward zero.

Figures \ref{Fig4}(a) and (b) display the numerically computed fractal dimensions of states on the ($E,V$) plane for $W<t$ and $W\geqslant t$, respectively. The localization properties differ in the two regimes, but the exact MEs agree well with the numerical results. For $W<t$, the system remains in the extended phase when $V<V_c$(with $V_c>W$). Because of the sign-changing divergence in $E^-_3$, states lying between (outside) two MEs are localized (extended) for $V_c\leqslant V<t$, while the opposite holds for $V>t$. The localization phase diagram is simpler for $W\geqslant t$: when $V\leqslant W$ the system is in the extended phase, whereas for $V>W$ a mixed phase appears in which states between (outside) two MEs are extended (localized).

Finally, we analyze how the MEs evolve on the $(E,V)$ plane as the parameters are varied. Equation (\ref{ME3}) shows that $t_1$ simply rescales the ME values. In Fig.\ref{Fig4}(c) we show MEs for different $W$ at a fixed $\phi$. For $W<t$, $E^-_3$ exhibits a sign-changing divergence at $V=t$, while $E^+_3$ remains continuous. As $W$ increases, their coincidence point moves toward ($V=t,E=-\infty$), and the MEs on the lower half plane shift to the right, delaying the emergence of mixed phase to larger $V$ values. On the upper half plane, the ME is essentially independent of $W$ when $W<t$. For $W\geqslant t$, two MEs are disconnected: $E^-_3$ lies on the upper half plane and $E^+_3$ on the lower half, both having $E=0$ and $V=W$ as horizontal and vertical asymptotes, respectively. Increasing $W$ shifts the vertical asymptote $V=W$ and the MEs to the right, while their overall structure  remains unchanged. In Fig.\ref{Fig4}(d) and its inset we present exact MEs for $W<t$ and $W>t$ at different values of $\phi$. For $W<t$, the $\phi$-dependence of MEs resembles that of the $W=0$ case [Fig.\ref{Fig2}(b)], except that the coincidence point satisfies $V_c>W$. As $\phi$ increases, the MEs on the lower half plane shift to the right, although parts of them always lie to the left of the divergence line $V = t$ (except at $\phi=\pi/2$), and the coincidence point moves to the lower right. On the upper half plane, the ME shifts to the left but always stays to the right of the divergence line. For $W\geqslant t$, the MEs depend only weakly on $\phi$: as $\phi$ increases, two MEs move slightly apart and the energy window widens a little.

\section{V. Conclusion and discussion}

We have investigated the localization phenomena and exact MEs in a quasiperiodic AB chain, which combines off-diagonal and diagonal AAH modulations with a synthetic gauge field. By analytically computing the LE using Avila's global theory, we have derived exact expressions for MEs separating extended, critical, and localized states. In the diagonal limit ($W=0$), the coupling to the $b$ sublattice generates hyperbolic MEs whose shapes can be continuously tuned by the magnetic flux $\phi$; these MEs undergo a sign-changing divergence at $V=t$. In the off-diagonal regime ($W \neq 0$), the interplay between quasiperiodic hopping and indirect tunneling through the $b$ sites gives rise to AMEs. Notably, critical states can emerge even when the hopping modulation $t_n$ does not possess incommensurately distributed zeros---a behavior distinct from traditional off-diagonal AAH models. For $\phi \neq 0$, the critical states are suppressed, and the MEs shift with the synthetic flux, thereby moving the phase boundaries. This parameter tunability offers a powerful experimental knob for engineering precise localization transitions in quasiperiodic AB chains.

Before discussing experimental realizations, we briefly clarify the relation between our model and related works. Our system indeed belongs to the general framework of quasiperiodic network models introduced in Ref.\cite{Hu2025}. It consists of a quasiperiodically modulated $a$-sublattice coupled to a periodic network ($b$-sublattice), and integrating out the network yields an energy-dependent effective Hamiltonian. However, the models in that work are diagonal, where the quasiperiodic modulation appears solely in the potentials, and MEs are derived via the self-duality of the effective classical AAH Hamiltonian. Here, we study the off-diagonal model, and beyond the duality-based arguments, we provide a rigorous analytical computation of the full LE via Avila's global theory. The novelty lies not in the mathematical framework itself, but in the new physical phenomena emerging in the side-coupled AB chain with a synthetic flux, most notably, flux-controlled AMEs and critical states absent in previous network models. In particular, the critical states originate from interference-induced zeros in the effective hopping, a mechanism distinct from those in other AAH scenarios. Thus, while our model shares the same methodological spirit as Ref.\cite{Hu2025}, the physics and analytical rigor presented here are substantially new. Additionally, flux-dependent MEs have also been studied in disordered zigzag chains \cite{An2018}, but those works focused on random disorder and did not address critical states or AMEs. Our AB chain provides exact analytical results for both conventional and anomalous MEs, where the flux serves as a continuous switch for criticality.

The single-particle physics and exact MEs predicted here can be experimentally realized and probed in many platforms, such as photonic waveguides \cite{Chang2025}, ultracold atoms \cite{Wang2022,Luschen2018}, cavity-polariton devices \cite{Goblot2020}, and superconducting processors \cite{Li2023}. Superconducting processors are particularly well-suited for realizing our model: the AB cage geometry can be directly mapped onto a qubit lattice with engineered nearest-neighbor couplings, while the synthetic magnetic flux can be engineered by applying continuous modulation tones to qubits, as demonstrated in recent experiments \cite{Rosen2024}. The diagonal and off-diagonal AAH ingredients on the $a$ sublattice have already been realized in superconducting circuits \cite{Li2023}. The localization of states can be observed using state tomography, while the energy spectrum can be obtained by Fourier transforming the experimentally detected response function \cite{Shi2023}. Our work establishes a rigorous theoretical framework for engineering controllable (A)MEs in such superconducting processors, paving the way for experimental studies of disorder-driven quantum phase transitions and critical states.


\section{Acknowledgments} 

This work is supported by the Innovation Program for Quantum Science and Technology under grant No. 2023ZD0300400 and the National Natural Science Foundation of China under grant No.12274419 and 12134015.

\section{Appendix: Computation of Lyapunov exponent for $W\neq0$}
\setcounter{equation}{0}
\def\theequation{A\arabic{equation}}

As shown in Eq.(\ref{LE}) in the main text, the LE is given by
\begin{equation}\label{A1}
	\gamma=\left\{  
	\begin{aligned}
		&\max(f_1+\gamma_C,0),\ \ \ V\leqslant W,\\  
		&\max(f_2+\gamma_C,0),\ \ \ V>W,
	\end{aligned} 
	\right.
\end{equation}
where
\begin{eqnarray}\label{A2}
	f_1&=&\ln |W/2|,\nonumber\\
	f_2&=&\ln|(V+\sqrt{V^2-W^2})/2|,\nonumber\\
	\gamma_C&=&-\frac{1}{2\pi}\int_0^{2\pi}\ln|t+W\cos(\xi)+t_1^2e^{i\phi}/E|\mathrm{d}\xi.
\end{eqnarray}
According to Jensen's formula \cite{Jen}, the integral in Eq.(\ref{A2}) is calculated as 
\begin{eqnarray}\label{A3}
	\gamma_C&=&\ln\left|\min\{|z_+|,|z_-|\}\right|-\ln |W/2|,\nonumber\\
	z_{\pm}&=&-\eta\pm\sqrt{\eta^2-1},\nonumber\\
	\eta&=&t/W+t_1^2e^{i\phi}/WE,
\end{eqnarray}
where $z_{\pm}$ are two roots of the equation $z^2+2\eta z+1=0$. Note that $z_+ z_- = 1$, so the two roots are reciprocals of each other. In the following we explicitly compute the LE by analyzing the moduli of two roots $|z_{\pm}|$. 

\subsection{A. Case $\phi=0$}

When $\phi=0$, the quantity $\eta=t/W+t_1^2/WE$ is real. When $|\eta|\leqslant1$ (or $\eta^2-1\leqslant0$), the moduli of two roots satisfy $|z_+|=|z_-|=1$, resulting in $\gamma_C=-\ln |W/2|$. Substituting the expression of $\eta$ in Eq.(\ref{A3}) into the inequality $|\eta|\leqslant1$ and performing a straightforward calculation, we obtain the parameter regimes satisfying $|\eta|\leqslant1$.
\begin{eqnarray}\label{A4}
	\left\{  
	\begin{aligned}
		&E\in\left[-\frac{t_1^2}{t-W},-\frac{t_1^2}{t+W}\right],\quad\quad W<t\\  
		&E\in\left(-\infty,-\frac{t_1^2}{t+W}\right]\cup\left[-\frac{t_1^2}{t-W},+\infty\right),W\geqslant t
	\end{aligned} 
	\right.
\end{eqnarray}
Substituting $|z_\pm|=1$ into $\gamma_C$, and then into $\gamma$ in Eq.(\ref{A1}), the LEs of states are expressed as
\begin{table}[H]
	\centering
	\begin{tabular}{|c|c|}
		\hline
	 	regime & LE $\gamma$ \\
		\hline	
		\thead{$W<t$, $V\leqslant W$, \\ $E\in\left[-\frac{t_1^2}{t-W},-\frac{t_1^2}{t+W}\right]$}
		&0\\
		\hline	
		\thead{$W\geqslant t$, $V\leqslant W$, \\ $E\in\left(-\infty,-\frac{t_1^2}{t+W}\right]\cup\left[-\frac{t_1^2}{t-W},+\infty\right)$}
		&0\\
		\hline	
		\thead{$W<t$, $V>W$, \\ $E\in\left[-\frac{t_1^2}{t-W},-\frac{t_1^2}{t+W}\right]$}
		&$\ln\left|\frac{V+\sqrt{V^2-W^2}}{W}\right|$\\
		\hline	
			\thead{$W\geqslant t$, $V>W$, \\ $E\in\left(-\infty,-\frac{t_1^2}{t+W}\right]\cup\left[-\frac{t_1^2}{t-W},+\infty\right)$}
		&$\ln\left|\frac{V+\sqrt{V^2-W^2}}{W}\right|$\\
		\hline	
	\end{tabular}.
\end{table}
Note that in the above cases, states are critical when $V\leqslant W$, since $f_1+\gamma_C$ is always zero. Furthermore, the finite LEs are energy-independent, which is highly unusual in systems with MEs. On the other hand, in the complementary regimes of Eq.(\ref{A4}), i.e.,
\begin{eqnarray}\label{A6}
	\left\{  
	\begin{aligned}
		&E\in\left(-\infty,-\frac{t_1^2}{t-W}\right)\cup\left(-\frac{t_1^2}{t+W},0\right)\cup\left(0,+\infty\right)\\
		&\quad\quad\quad\quad\quad\quad\quad\quad\quad\quad\quad\quad\quad\quad\quad\quad W<t,  \\ 
		&E\in\left(-\frac{t_1^2}{t+W},0\right)\cup\left(0,-\frac{t_1^2}{t-W}\right), W\geqslant t,
	\end{aligned} 
	\right.
\end{eqnarray}
we have $|\eta|>1$, and the moduli of two real roots are no longer equal. For $W<t$, two roots satisfy $|z_-|>1>|z_+|$  in the energy interval $\left(-\infty, -t_1^2/(t-W)\right)\cup\left(0,+\infty\right)$, while $|z_+|>1>|z_-|$ in the energy interval $\left(-t_1^2/(t+W),0\right)$. For $W\geqslant t$, $|z_+|>1>|z_-|$ holds in the energy interval $\left(-t_1^2/(t+W),0\right)$, while $|z_-|>1>|z_+|$ in the energy interval $\left(0,-t_1^2/(t-W)\right)$. From Eqs.(\ref{A3}-\ref{A1}), the LEs of states are expressed as
\begin{table}[H]
	\centering
	\begin{tabular}{|c|c|}
		\hline
		regime & LE $\gamma$ \\
		\hline	
		\thead{$W<t$, $V\leqslant W$, \\ $E\in\left(-\infty,-\frac{t_1^2}{t-W}\right)\cup\left(0,+\infty\right)$}
		&$\max\{\ln|z_+|,0\}=0$\\
		\hline	
		\thead{$W<t$, $V\leqslant W$, \\ $E\in\left(-\frac{t_1^2}{t+W},0\right)$}
		&$\max\{\ln|z_-|,0\}=0$\\
		\hline	
		\thead{$W\geqslant t$, $V\leqslant W$, \\ $E\in(0,-\frac{t_1^2}{t-W})$}
		&$\max\{\ln|z_+|,0\}=0$\\
		\hline	
		\thead{$W\geqslant t$, $V\leqslant W$, \\ $E\in\left(-\frac{t_1^2}{t+W},0\right)$}
		&$\max\{\ln|z_-|,0\}=0$\\
		\hline	
		\thead{$W<t$, $V> W$, \\ $E\in\left(-\infty,-\frac{t_1^2}{t-W}\right)\cup\left(0,+\infty\right)$}
		&$\max\left\{\ln\left|\frac{V+\sqrt{V^2-W^2}}{W}z_+\right|,0\right\}$\\
		\hline	
		\thead{$W<t$, $V> W$, \\ $E\in\left(-\frac{t_1^2}{t+W},0\right)$}
		&$\max\left\{\ln\left|\frac{V+\sqrt{V^2-W^2}}{W}z_-\right|,0\right\}$\\
		\hline	
		\thead{$W\geqslant t$, $V> W$, \\ $E\in(0,-\frac{t_1^2}{t-W})$}
		&$\max\left\{\ln\left|\frac{V+\sqrt{V^2-W^2}}{W}z_+\right|,0\right\}$\\
		\hline	
		\thead{$W\geqslant t$, $V> W$, \\ $E\in\left(-\frac{t_1^2}{t+W},0\right)$}
		&$\max\left\{\ln\left|\frac{V+\sqrt{V^2-W^2}}{W}z_-\right|,0\right\}$\\
		\hline	
	\end{tabular},
\end{table}
where
\begin{equation}
	z_\pm=-\frac{t}{W}-\frac{t_1^2}{WE}\pm\sqrt{\left(\frac{t}{W}+\frac{t_1^2}{WE}\right)^2-1}.
\end{equation}
Note that in the above cases, states are extended when $V\leqslant W$, since $f_1+\gamma_C<0$. Combining the above expressions for LEs, we obtain the following information. When $V\leqslant W$, the exact expressions of AMEs, separating extended and critical states, are
\begin{equation}
	E^\pm_1=-\frac{t_1^2}{t\pm W}.
\end{equation}
When $V>W$, the MEs, separating extended and localized states, are given by
\begin{eqnarray}
	E^\pm_2=\frac{t_1^2}{\pm V-t},
\end{eqnarray} 
which are determined by the condition $\left|(V+\sqrt{V^2-W^2})z_\pm/W\right|=1$.

\subsection{B. Case $\phi\neq0$}
When $\phi\neq0$, the quantity $\eta=t/W+t_1^2e^{i\phi}/WE$ is complex.  One can easily prove that 
\begin{eqnarray}\label{A14}
	\left\{  
	\begin{aligned}
		&	\mathrm{Re}(\eta)>0\Rightarrow|z_-|>1>|z_+|,\\
		&\mathrm{Re}(\eta)\leqslant0\Rightarrow|z_+|>1>|z_-|.
	\end{aligned} 
	\right.
\end{eqnarray}
where $\mathrm{Re}(\eta)$ is the real part of $\eta$. The condition $\mathrm{Re}(\eta)>0$ is satisfied when $E\in\left(-\infty,-t_1^2\cos\phi/t\right)\cup\left(0,+\infty\right)$, while $E\in\left[-t_1^2\cos\phi/t,0\right)$ corresponds to $\mathrm{Re}(\eta)\leqslant0$. Returning to Eqs.(\ref{A3}-\ref{A1}), the LEs of states are expressed as
\begin{table}[H]
	\centering
	\begin{tabular}{|c|c|}
		\hline
		regime & LE $\gamma$ \\
		\hline	
		\thead{$V\leqslant W$, \\ $E\in\left(-\infty,-\frac{t_1^2\cos\phi}{t}\right)\cup\left(0,+\infty\right)$}
		&$\max\{\ln|z_+|,0\}=0$\\
		\hline	
		\thead{$V\leqslant W$, \\ $E\in\left[-\frac{t_1^2\cos\phi}{t},0\right)$}
		&$\max\{\ln|z_-|,0\}=0$\\
		\hline	
		\thead{$V>W$, \\ $E\in\left(-\infty,-\frac{t_1^2\cos\phi}{t}\right)\cup\left(0,+\infty\right)$}
		&$\max\left\{\ln\left|\frac{V+\sqrt{V^2-W^2}}{W}z_+\right|,0\right\}$\\
		\hline	
		\thead{$V>W$, \\ $E\in\left[-\frac{t_1^2\cos\phi}{t},0\right)$}
		&$\max\left\{\ln\left|\frac{V+\sqrt{V^2-W^2}}{W}z_-\right|,0\right\}$\\
		\hline	
	\end{tabular}.
\end{table}
The MEs separating extended and localized states are determined by the equation
\begin{eqnarray}
	\left|\frac{\left(V+\sqrt{V^2-W^2}\right)}{W\max|z_{\pm}|}\right|=1,
\end{eqnarray}
which results in
\begin{eqnarray}
	E_3^\pm=\frac{-t_1^2\left[\left(\frac{S^2+W^2}{S^2-W^2}\right)^2\sin^2\phi+\cos^2\phi\right]}{t\cos\phi\pm V\sqrt{1-\frac{4S^2(t^2-W^2)}{(S^2-W^2)^2}\sin^2\phi}},
\end{eqnarray}
where $S=V+\sqrt{V^2-W^2}$. Notice that $E_3^\pm$ are physically meaningful only when the expression under the square root is non-negative, which holds when $W\geqslant t$. However, when $W<t$, two MEs are complex in the interval $W<V<V_c$ and become real in the interval $V\geqslant V_c$, where 
\begin{eqnarray}
	V_c&=&\frac{G^2+W^2}{2G},\nonumber\\
	G&=&\sqrt{2(t^2-W^2)\sin^2\phi+W^2+2\sin\phi\sqrt{M}},\nonumber\\
	M&=&(t^2-W^2)(W^2\cos^2\phi+t^2\sin^2\phi).
\end{eqnarray}

\end{document}